\documentclass[10pt]{iopart}

\usepackage{latexsym,epsfig,graphicx}
\usepackage{cite}

\newcommand{\ket}[1]{\vert #1\rangle}
\newcommand{\bra}[1]{\langle #1\vert}

\newcommand{\ii}{\mathrm{i}}
\newcommand{\dd}{\mathrm{d}}
\newcommand{\qq}{\mathbf{q}}
\newcommand{\eh}{\mathrm{e}}
\renewcommand{\Tr}{\mathrm{Tr}}

\renewcommand{\br}{\mathbf{r}}

\newcommand{\diff}{\mathrm{d}}

\newcommand{\an}[1]{\hat{#1}}
\newcommand{\cre}[1]{\hat{#1}^\dag}

\begin{document}

\title{Dynamics, dephasing and clustering of impurity atoms in Bose-Einstein condensates}

\author{Alexander Klein$^{1,2}$, Martin Bruderer$^{1}$,
Stephen R. Clark$^{1,3}$, and Dieter Jaksch$^{1,2}$}

\address{$^1$Clarendon Laboratory, University of Oxford,
Parks Road, Oxford OX1 3PU, United Kingdom \\
$^2$ Keble College, Parks Road, Oxford OX1 3PG, United Kingdom \\
$^3$ Trinity College, Broad Street, Oxford OX1 3BH, United Kingdom
}

\begin{abstract}

We investigate the influence of a Bose-Einstein condensate (BEC) on
the properties of immersed impurity atoms, which are trapped in an
optical lattice. Assuming a weak coupling of the impurity atoms to
the BEC, we derive a quantum master equation for the lattice system.
In the special case of fixed impurities with two internal states the
atoms represent a quantum register and the quantum master equation
reproduces the exact evolution of the qubits. We characterise the
qubit dephasing which is caused by the interspecies coupling and
show that the effect of sub- and superdecoherence is observable for
realistic experimental parameters. Furthermore, the BEC phonons
mediate an attractive interaction between the impurities, which has
an important impact on their spatial distribution. If the lattice
atoms are allowed to move, there occurs a sharp transition with the
impurities aggregating in a macroscopic cluster at experimentally
achievable temperatures. We also investigate the impact of the BEC
on the transport properties of the impurity atoms and show that a
crossover from coherent to diffusive behaviour occurs with
increasing interaction strength.

\end{abstract}

\date{\today}

\maketitle




\section{Introduction}

Ultracold atoms in optical lattices have attracted considerable
interest during the last few years. Theoretical investigations
showed that ultracold atoms in optical lattices can be used for
mimicking a wide range of models encountered in condensed matter
physics \cite{Lewenstein-AiP-2006}. These models include the Hubbard
Hamiltonian \cite{Jaksch-PRL-1998,Jaksch-Ann-2005}, spin-spin
interactions \cite{Sorensen-PRL-1999,Duan-PRL-2003},
high-temperature superconductivity
\cite{Hofstetter-PRL-2002,Klein-PRA-2006b}, effective magnetic
fields
\cite{Jaksch-NJP-2003,Juzeliunas-PRL-2004,Juzeliunas-PRA-2005,
Juzeliunas-PRA-2006,Mueller-PRA-2004,Sorensen-PRL-2005}, even with
non-abelian gauge potentials
\cite{Osterloh-PRL-2005,Ruseckas-PRL-2005}, and the fractional
quantum Hall effect \cite{Sorensen-PRL-2005,Palmer-PRL-2006}, to
name but a few. Experimental efforts have led to an unprecedented
control over the properties of optical lattice systems
\cite{Bloch-NatPhys-2005}, including such milestones as the Mott
insulator to superfluid transition \cite{Greiner-Nature-2002,
Greiner-Nature-2002b}, investigations of Bose-Fermi
\cite{Guenter-PRL-2006,Ospelkaus-PRL-2006} as well as Bose-Bose
mixtures \cite{Catani-2007}, vortex pinning \cite{Tung-PRL-2006},
the creation of repulsively bound atom pairs
\cite{Winkler-nature-2006}, and cold controlled collisions between
atoms in optical lattices \cite{Mandel-Nature-2003,
Mandel-PRL-2003}. Especially for the latter experiments the absence
of lattice phonons and thus the suppression of decoherence
mechanisms was crucial. However, when mimicking the behaviour of
electrons in crystals this lack of phonons might lead to an
oversimplification of the underlying model and it is desirable to
introduce phonons in a controlled manner into the optical lattice
system.

One way of achieving this goal is to immerse the optical lattice
system into a Bose-Einstein condensate (BEC). Experiments where both
atom species are trapped by the optical lattice are common and
decoherence effects in such systems have already been observed
\cite{Guenter-PRL-2006,Ospelkaus-PRL-2006}. Furthermore, these
Bose-Fermi mixtures promise rich phase diagrams including charge and
spin density wave phases \cite{Mathey-PRL-2004,Pazy-PRA-2005},
pairing of fermions with bosons \cite{Lewenstein-PRL-2004} and a
supersolid phase \cite{Buchler-PRL-2003}. We, however, focus on the
case where only one species is trapped by the optical lattice. This
can be achieved by a suitable choice of the laser wavelengths and
atomic species such that the BEC is not affected by the optical
lattice \cite{Leblanc-PRA-2007}. The lattice atoms interact with the
condensate via density-density interaction, which can be described
by Bogoliubov phonons coupling to the impurities. Earlier studies on
such systems have shown that this coupling can be exploited to cool
the lattice atoms to extremely low temperatures
\cite{Griessner-PRL-2006,Griessner-NJP-2007}, which are otherwise
very difficult to achieve. In Ref.~\cite{Bruderer-PRA-2007}, the
present authors have derived a model in which lattice atoms dressed
by a coherent state of Bogoliubov phonons constitute polarons
\cite{Mahan-2000}. The model exhibits an attractive interaction
potential between the lattice atoms
\cite{Bardeen-PR-1967,Klein-PRA-2005} and allows a generalised
master equation to be deduced which shows that the system exhibits a
crossover from coherent to diffusive dynamics.

Instead, in this work we concentrate on describing the lattice atoms
by a quantum master equation (QME), which is derived in
Sec.~\ref{Sec:QME}. We show in Sec.~\ref{Sec:QuantReg} that for the
case of fixed impurities the system can be solved exactly and after
tracing out the BEC degrees of freedom the QME reproduces this exact
solution. Due to the coupling to the phonons the lattice atoms
experience dephasing, which can be used to demonstrate the effects
of sub- and superdecoherence \cite{Palma-Proc-1996} and to probe
spatial properties of the BEC analogous to
Ref.~\cite{Bruderer-NJP-2006}. We also investigate the severe effect
this decoherence has if the atoms in the lattice are used as a
quantum register. In Sec.~\ref{Sec:Cluster} we show that the
attractive interaction mediated between the lattice atoms leads to
the formation of atom clusters, which should be observable for
typical experimental parameter regimes. This effect is reminiscent
of the clustering of ad-atoms on crystal surfaces
\cite{Jensen-RMP-1999}. The influence of the phonon coupling on the
transport properties of the lattice atoms is investigated in
Sec.~\ref{Sec:QTrans}. We show that the crossover from coherent to
diffusive transport, which was observed in
Ref.~\cite{Bruderer-PRA-2007}, can also be described by the QME. In
contrast to the treatment in Ref.~\cite{Bruderer-PRA-2007}, the QME
in addition gives access to the off-diagonal elements of the density
operator. We also show the limitations of the QME by applying it to
a tilted lattice system and comparing the results to a near-exact
numerical solution for the time evolution.


\section{Model\label{Sec:Model}}

\begin{figure}
\begin{center}
  \includegraphics[width=9cm]{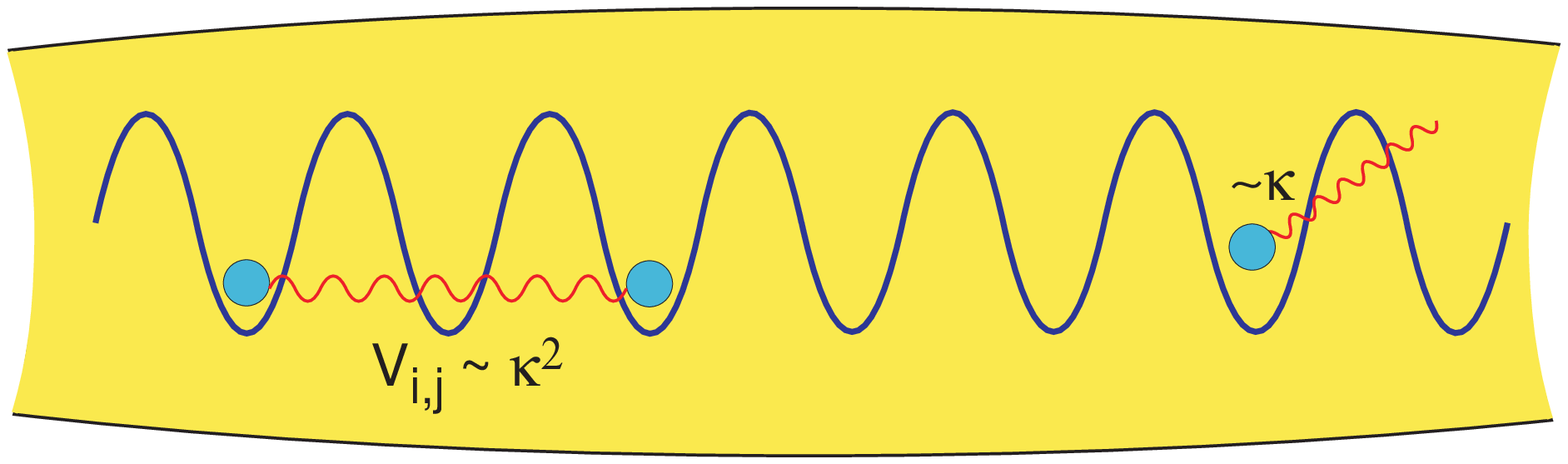}
  \caption{An ultracold quantum gas trapped in an optical lattice is
  immersed in a much larger BEC. The impurities interact with the BEC
  via the coupling constant $\kappa$ and can excite Bogoliubov
  phonons (right). These lead to dephasing effects.
  By exchanging phonons an off-site interaction $V_{i,j}$
  is mediated between the lattice atoms (left).
  \label{Fig:Scheme}}
\end{center}
\end{figure}

The system under consideration is composed of a BEC and an ultracold
gas of impurity atoms trapped in an optical lattice giving a setup
like that shown in Fig.~\ref{Fig:Scheme}. The impurities can be
either bosonic or fermionic atoms. For most of the results derived
in this work the statistics of the lattice atoms is unimportant, but
for concreteness we will assume bosonic impurities in the following.
The dynamics of the whole system is governed by the Hamiltonian
$\hat{H} = \hat{H}_{\mathrm{L}} + \hat{H}_{\mathrm{B}} +
\hat{H}_{\mathrm{I}}$, where
\begin{equation}
   \hat{H}_{\mathrm{B}} = \int\dd \br\, \cre{\phi}(\br)\left[-\frac{\hbar^2\nabla^2}{2
    m_b}+V_{\mathrm{ext}}(\br)+\frac{g}{2}\cre{\phi}(\br)\an{\phi}(\br)\right]\an{\phi}(\br)
\end{equation}
describes the BEC, with $\cre{\phi}$ the condensate field operator,
$m_b$ the mass of the BEC atoms, $V_{\mathrm{ext}}(\br)$ an external
potential confining the BEC, and $g>0$ the coupling constant between
the BEC atoms. Here, we have assumed that the optical lattice
potential does not affect the BEC atoms, which can be achieved by
choosing the laser wavelengths and atom species accordingly
\cite{Leblanc-PRA-2007}. The atoms in the optical lattice, which are
distinguishable from the atoms in the BEC, are described by the
Hamiltonian $\hat{H}_{\mathrm{L}}$, whereas the interaction between
the two sub-systems is given by
\begin{equation}
\hat{H}_{\mathrm{I}} = \kappa\int
    \dd\br\,\cre{\chi}(\br)\an{\chi}(\br)\cre{\phi}(\br)\an{\phi}(\br)\,.
\end{equation}
Here, $\cre{\chi}$ is the field operator of the lattice atoms, and
$\kappa$ is the density-density coupling constant between the BEC
and the impurities. It was shown in Ref.~\cite{Bruderer-PRA-2007}
that in the tight-binding limit and under the condition $|\kappa|/g
n_0\xi^D\ll1$ (where $D$ is the dimensionality of the system, $n_0$
the density of the BEC in the trap centre, and $\xi =
\hbar/\sqrt{m_b g n_0}$ the healing length) the total Hamiltonian
reduces to a Hubbard-Holstein model, given by
\begin{eqnarray}
  \hat H_\mathrm{L} = - J \sum_{\langle i,j \rangle} \hat
  a_j^\dagger \hat a_i + \frac{U}{2} \sum_i \hat n_i (\hat n_i-1) +
   \mu \sum_i  \hat n_i \,, \\ \label{Eq:H_I}
   \hat H_\mathrm{I} = \left.\sum_{\nu}\right.^\prime \sum_j \hbar\omega_\nu \left[M_{j,\nu}\an{b}_\nu +
    M^\ast_{j,\nu}\cre{b}_\nu\right]\an{n}_j \,,\\ \label{Eq:H_BEC}
  \hat H_\mathrm{BEC} =  \left.\sum_{\nu}\right.^\prime \hbar\omega_\nu\cre{b}_\nu\an{b}_\nu
  \,.
\end{eqnarray}
Here, the prime at the sum indicates that zero energy modes have
been excluded, $\cre{b}_\nu$ creates a Bogolibov phonon in mode
$\nu$ with energy $\hbar \omega_\nu$, $\cre{a}_j$ creates a lattice
atom in site $j$, $\hat n_j = \cre{a}_j \an{a}_j$ is the number
operator on site $j$, $J$ describes the hopping of the impurities,
$U$ is their on-site interaction, and the coupling between the
phonons and the lattice atoms is given by
\begin{equation}
  F_{j,\nu} = \hbar \omega_\nu M_{j,\nu} = \kappa \int \! \dd \br \, \phi_0(\br) \left[u_\nu(\br)
- v_\nu(\br)\right]|\eta_j(\br)|^2 \,.
\end{equation}
In this equation, $\eta_j$ is a Wannier function describing an atom
in lattice site $j$, $\phi_0$ is the solution of the
Gross-Pitaevskii equation and $u_\nu$ and $v_\nu$ solve the
Bogoliubov-de\,Gennes equations, see also
Ref.~\cite{Oehberg-PRA-1997}.

For the case of a homogenous condensate, which we consider in the
remainder of the paper, the expressions for the atom-phonon coupling
is given by
\begin{equation}
 F_{j,\qq} = \kappa \sqrt{\frac{ n_0
    \varepsilon_{\qq}}{\hbar\omega_{\qq}}} \, f_j(\qq) \,,
\end{equation}
where $\qq$ is the phonon quasi-momentum, $\varepsilon_{\qq} =
(\hbar \mathbf{q})^2/2m_b$ the free particle energy,
$\hbar\omega_\qq= \sqrt{\varepsilon_{\qq}
(\varepsilon_{\mathbf{q}}+2g n_0)}$ the Bogoliubov dispersion
relation and $f_j(\qq)=\Omega^{-1/2}\int \dd \mathbf{r}
|\eta_j(\br)|^2\exp(\ii \qq\cdot\br)$, with $\Omega$ the
quantisation volume. The latter integral can in general not be
solved analytically. However, for sufficiently deep lattices
\cite{Bloch-2007}, the Wannier functions can be approximated by
Gaussians, yielding
\begin{equation}
f_j(\qq) = \frac{\eh^{\ii \qq \cdot \br_j}}{\sqrt{\Omega}}
\exp\left(- \frac{1}{4} \sum_{l = 1}^D q_l^2 x_0^2\right) \,,
\end{equation}
with $x_0 = \sqrt{\hbar / m_b \omega_\mathrm{t}}$, where
$\omega_\mathrm{t}$ is the trapping frequency of a harmonic trap
approximating the lattice potential at a given lattice site.

An experimental realisation of this setup is achievable with present
techniques as follows. The creation of Rb condensates with $10^{6}$
atoms has been demonstrated leading to the desired BEC densities of
about $10^{20}/\mathrm{m}^3$ in three dimensions, see for instance
\cite{Theis-PRL-2004}. By choosing a sufficiently flat trapping
potential of a few Hz the BEC can be assumed to be homogenous to a
good approximation in the centre of the trap, extending over a few
micrometers. Furthermore, in references
\cite{Guenter-PRL-2006,Ospelkaus-PRL-2006} a mixture of Rb and K
atoms has been created and trapped in a three-dimensional optical
lattice, where several tens of thousands of lattice sites have been
occupied. The filling for the K atoms ranged between 0 and 1 atoms
per lattice site, whereas the filling of the Rb atoms could exceed 5
atoms per site in the centre of the trap. Although in these
experiments both atom species were trapped by the optical lattice,
techniques to trap only one of the two species, for example K, have
been studied extensively in reference \cite{Leblanc-PRA-2007}.
Applying these techniques leaves the Rb atoms virtually unaffected
and enables the creation of a nearly homogenous BEC.


\section{The quantum master equation \label{Sec:QME}}

In order to investigate the behaviour of the impurity atoms we
derive a Quantum Master Equation (QME) for the lattice system by
tracing out the surrounding BEC. The details are given in
\ref{App:QME}. Here we note that the main condition of deriving the
QME is that the sound velocity of the condensate, $ c \sim \sqrt{g
n_0/m_b}\,$, is larger than the typical hopping speed of the atoms,
i.e., $c \gg J a/\hbar$, where $a = \lambda /2$ is the distance
between two lattice sites. With this assumption we get
\begin{eqnarray} \label{Eq:QME}
  \nonumber \fl
\ii \hbar \partial_t \hat \varrho_L (t) = \left[
  \hat H_g (t) \,,\, \hat \varrho_L (t) \right] &
  - \frac{\ii}{\hbar} \left.\sum_\qq\right.^\prime \sum_{l,l'}
  \frac{\sin(\omega_\qq t)}{\omega_\qq}
  \left(\hat n_l \hat n_{l'} \hat \varrho_L(t) +
   \hat \varrho_L(t)\hat n_l \hat n_{l'} -2  \hat n_{l'}
    \hat \varrho_L(t) \hat n_{l} \right) \\ \fl
    &\qquad
   \times \left( F_{\qq,l}F_{\qq,l'}^\ast + N_\qq(T) (F_{\qq,l} F_{\qq,l'}^\ast +
   F^\ast_{\qq,l} F_{\qq,l'})\right) \,.
\end{eqnarray}
Here, $N_\qq(T) = 1/(\exp(\hbar \omega_\qq / k_\mathrm{B} T)-1)$ is
the number of thermal phonons at temperature $T$, and $\hat
\varrho_\mathrm{L}(t) = \Tr_\mathrm{B} \hat \varrho (t)$, where
$\Tr_\mathrm{B}$ denotes the trace over the condensate. Furthermore,
the Hamiltonian
\begin{equation}
 \label{Eq:H_total_red}\fl
  \hat H_g (t) = \hat H_L - \left.\sum_\qq\right.^\prime \sum_{l,l'}
  \frac{1 - \cos (\omega_\qq t)}{2 \hbar
  \omega_\qq}
  (F_{\qq,l} F_{\qq,l'}^\ast +
  F_{\qq,l}^\ast F_{\qq,l'})\hat n_l(t) \hat n_{l'} (t)
\end{equation}
describes the coherent evolution of the lattice atoms, most notably
an off-site interaction which is mediated by the phonons of the BEC.
The interaction term includes a transient behaviour described by the
cosine functions, which accounts for suddenly turning on the
interaction between lattice atoms and BEC. The sum over all these
cosine functions vanishes in the limit $t \to \infty$.

The Born approximation used in deriving the QME is valid if the
perturbation caused by the BEC is small compared to the typical
energy scales given by the lattice system. If only the hopping term
is important for the lattice atoms, this energy scale is determined
by $J$. The energy scale of the perturbation is given by the
so-called polaron energy $E_p = \sum_{\qq}^\prime (F_{\qq,l}
F_{\qq,l}^\ast + F_{\qq,l}^\ast F_{\qq,l})/2 \hbar \omega_\qq \sim
\kappa^2/g \xi^D $, which leads to the condition $J \gg E_p$. This
parameter regime is complementary to the one considered in
Ref.~\cite{Bruderer-PRA-2007}, where the opposite limit was
investigated. As we will show in the following section, the
condition $J \gg E_p$ is not required for the case of $J=0$, where
an analytical solution of the dynamics of the whole system can be
derived. After tracing out the BEC degrees of freedom this solution
agrees with the one given by the QME.


\section{Fixed impurities and the Quantum Register \label{Sec:QuantReg}}

For very deep optical lattices the hopping constant $J$ is
essentially zero and the impurities cannot leave their site. If
there is a maximum of one atom in each lattice site, the setup can
be used as a quantum register \cite{Schrader-PRL-2004}. The lattice
atoms represent qubits on which single qubit rotations can be
implemented via external laser pulses. For the manipulation of the
atoms single site addressability is necessary, which may be achieved
by using infrared lattices \cite{Scheunemann-PRA-2000}, by leaving
empty sites between the atoms
\cite{Miroshnychenko-Nature-2006,Miroshnychenko-NJP-2006}, by
exploiting the properties of marker atoms \cite{Calarco-PRA-2004},
or by additional external fields
\cite{Zhang-PRA-2006,Gorshkov-2007}. An entangling two-qubit gate
can be implemented by using the interaction which is mediated by the
condensate \cite{Klein-PRA-2005}. For this it is necessary to turn
the interaction on and off, which can be done by using different
internal states of the lattice atoms, some of which couple to the
BEC, and others that do not couple. However, due to the coupling to
the BEC the qubits also experience dephasing, which will be
investigated in this section.

\subsection{Analytical solution of the time evolution \label{Sec:AnalytSol}}

We first calculate the time evolution of the atoms subject to the
BEC coupling. For easier notation we focus on the case where the
lattice atoms have only one internal state and we assume that there
is a maximum of one atom per lattice site. In the regime were our
model introduced in section \ref{Sec:Model} is valid,
i.e.~$|\kappa|/g n_0 \xi^D \ll 1$, and for $J=0$, the Hamiltonian of
the whole system simplifies to $\hat H_0 = \hat H_{\mathrm{BEC}} +
\hat H_{\mathrm{I}}$, where $\hat H_{\mathrm{I}}$ is given in
Eq.~(\ref{Eq:H_I}) and $\hat H_{\mathrm{BEC}}$ in
Eq.~(\ref{Eq:H_BEC}). The time evolution $\hat U(t) = \exp(-\ii \hat
H_0 t/\hbar)$ of this Hamiltonian is solved analytically. We first
note that $\hat U(t) = \prod_\qq^\prime \hat U_\qq(t)$ with
\begin{equation}
\hat U_\qq(t) = \exp \left[- \ii \left( \hbar \omega_\qq \hat
b^\dagger_\qq \hat b_\qq  +\sum_l  \hat n_l (F_{l,\qq} \hat b_\qq +
   F_{l,\qq}^\ast \hat b^\dagger_\qq    )\right) \frac{t}{\hbar} \right] \,.
\end{equation}
This operator can be decomposed using the methods described in
Ref.~\cite{Wei-JMP-1963}, yielding
\begin{eqnarray}\label{Eq:Exact_p1}
  \nonumber \fl
  \hat U_\qq(t) =& \exp\left[- \ii  \omega_\qq  \hat
b^\dagger_\qq \hat b_\qq t\right] \, \exp\left[\sum_l
  \hat n_l (\tilde F_{l,\qq} \hat b_\qq -
  \tilde F^\ast_{l,\qq} \hat b_\qq^\dagger)\right]
  \, \\ \fl
  & \times\exp\left[ \frac{\ii (\omega_\qq t - \sin(\omega_\qq t))}{2 \hbar^2 \omega_\qq^2}
   \sum_{l,l'} \hat n_l \hat n_{l'}
  (F_{l,\qq} F^\ast_{l',\qq} + F^\ast_{l,\qq} F_{l',\qq}) \right] \,.
\end{eqnarray}
Here, we have defined $\tilde F_{l,\qq} = F_{l,\qq} \left[- \ii
\sin(\omega_\qq t) - \left(1-\cos(\omega_\qq t)\right)\right]/\hbar
\omega_\qq$. The last term in Eq.~(\ref{Eq:Exact_p1}) describes the
off-site interaction between two atoms confined in lattice sites $l$
and $l'$, which is mediated by a phonon with quasi-momentum $\qq$.
The total interaction potential is derived by adding all phonon
contributions together. For $t \to \infty$, the oscillations caused
by the sine function cancel each other and the off-site interaction
potential is given by\footnote{For later convenience, we define this
$V_{l,l'}$ in such a way that the interaction term in the
Hamiltonian is given by $-\sum_{l,l'} V_{l,l'} \hat n_l \hat
n_{l'}$.}
\begin{equation} \label{Eq:Vlls}
 V_{l,l'} =
   \left.\sum_{\qq}\right.^\prime \frac{F_{l,\qq} F^\ast_{l',\qq} + F^\ast_{l,\qq}
   F_{l',\qq}}{ 2 \hbar \omega_\qq}  \,.
\end{equation}
This interaction potential was already found earlier
\cite{Klein-PRA-2005,Bruderer-PRA-2007}, however, the exact solution
in addition reveals the transient behaviour after suddenly turning
on the interaction, which is described by the sine functions.

The coupling to the BEC leads to dephasing of the lattice atoms as
gets apparent after tracing out the condensate degrees of freedom.
There are in essence two different types of dephasing. One can be
observed when a lattice atom is driven into a superposition of two
different states which couple differently to the condensate. This
situation has been discussed in reference \cite{Bruderer-NJP-2006}.
The other type of dephasing is given when comparing the phase of two
atoms in different lattice sites. We illustrate this effect of
dephasing by calculating the correlation function $ \left\langle
\hat a^\dagger_\gamma \hat a_\beta \right\rangle = \mathrm{Tr}\left[
\hat a^\dagger_\gamma \hat a_\beta \hat \varrho (t) \right]  =
\mathrm{Tr}\left[ \hat U^\dagger(t) \hat a^\dagger_\gamma \hat U(t)
\hat U^\dagger(t) \hat a_\beta \hat U(t) \hat \varrho (0) \right]$.
A combination of the two dephasing effects is discussed in section
\ref{Sec:Subandsuper}.

If we choose the same initial conditions as in~\ref{App:QME}, namely
$\hat \varrho(0) = \hat \varrho_\mathrm{L}(0) \otimes \hat
\varrho_\mathrm{B}(0)$ and $\hat \varrho_\mathrm{B}(0)$ describes a
thermal state of the BEC, we get
\begin{eqnarray}
  \nonumber \fl
  \left\langle \hat a^\dagger_\gamma \hat a_\beta \right\rangle = &
  \Tr_\mathrm{L} \left[   \exp\left( \frac{\ii}{\hbar} \int_0^t \hat H_g(s) \,\diff s\right)
  \tilde a^\dagger_\gamma(0) \tilde a_\beta(0)
  \exp\left( -\frac{\ii}{\hbar} \int_0^t \hat H_g(s) \,\diff
  s\right) \tilde \varrho_L(0) \right] \times \\ \fl
  &\quad
   \exp \left(- \left. \sum_\qq \right.^\prime \frac{1}{\hbar^2 \omega_\qq^2}
  | F_{\beta,\qq} - F_{\gamma,\qq}|^2 (1 - \cos(\omega_\qq t)) ( 2N_\qq(T) + 1)
  \right) \,.    \label{Eq:CorrFunDec}
\end{eqnarray}
Here, $\Tr_\mathrm{L}$ denotes the trace over the lattice system,
$\hat H_g$ was introduced in Eq.~(\ref{Eq:H_total_red}), and we made
use of the identity $\Tr_\mathrm{B} \left[\exp(\alpha_\qq \hat
b^\dagger_\qq - \alpha^\ast_\qq \hat b_\qq)  \hat
\varrho_\mathrm{B}\right] = \exp[- |\alpha_\qq|^2 (2 N_\qq(T)
+1)/2]$ for thermally distributed $\hat \varrho_\mathrm{B}$ and some
complex number $\alpha_\qq$ \cite{Breuer-2002}. The first part of
Eq.~(\ref{Eq:CorrFunDec}) describes the correlations between lattice
sites $\gamma$ and $\beta$ which are induced by the dynamics of the
lattice atoms. In our case, their time evolution only contributes a
phase term. The second part describes the dephasing which is caused
by the coupling to the BEC atoms. Let us denote the dephasing term
by $\Gamma$. In the thermodynamic limit of the Bogoliubov modes we
can replace the sum by an integral and get
\begin{eqnarray} \nonumber \fl
  \Gamma &\geq \exp\left( - \left. \sum_\qq \right.^\prime
  \frac{8 d_\qq}{(\hbar \omega_\qq)^2} (2N_\qq(T) +1) \right) \\ \fl
  &\approx
  \exp\left( - \int \! \diff^D \mathbf{k} \frac{8 \kappa^2}{(2 \pi)^D g^2 n_0 \xi^D}
   \frac{\exp(-\mathbf{k}^2 x_0^2/2 \xi^2)}{k \sqrt{k^2+2}^3} (2 N_\mathbf{k}(T) +1)
   \right) =: \Gamma_\mathrm{min} \,, \label{Eq:Gamma_est}
\end{eqnarray}
where $d_\qq = \kappa^2 n_0 \varepsilon_\qq  \exp(- \qq^2 x_0^2/2) /
\Omega \hbar \omega_\qq$, $F_{\qq,j} = \sqrt{d_\qq} \exp(\ii \qq
\br_j)$, and $D$ is the number of spatial dimensions. For zero
temperature, $D=3$, and $\xi \gg x_0$ the integral can be
approximated by neglecting the exponential function which gives
$\Gamma(T=0) \geq \exp( -4 \kappa^2 / \sqrt{2} \pi^2 g^2 n_0
\xi^3)$, whereas for sufficiently high finite temperature
$k_\mathrm{B} T \gg g n_0$ we find $\Gamma \geq \exp(- \kappa^2
k_\mathrm{B} T /\sqrt{2} \pi g^3 n_0^2 \xi^3)$. In both cases the
factor $\Gamma$ describing the dephasing fulfils $\Gamma > \varsigma
>0$ for a suitable real number $\varsigma$.
This is independent of time or of the distance between the two
atoms, which is different from the one-dimensional case, where
numerical calculations show that $\Gamma$ decays to 0 in the
thermodynamic limit for $t \to \infty$ as well as for $|r_\gamma -
r_\beta| \to \infty$, indicating that no correlations survive.

Interestingly, the solution of the QME introduced in
Sec.~\ref{Sec:QME} gives the same correlation functions, which also
holds for the time evolution of the operators $\hat a_j$ after
tracing out the condensate. This is due to the fact that the
interaction commutes with the lattice Hamiltonian for $J=0$ and that
the BEC is initially in a Gaussian, namely a thermal,
state~\cite{Doll-2007}. Hence, for the case $J=0$ the QME describes
the lattice atoms exactly after the trace over the BEC is taken. For
$J \neq 0$ the interaction Hamiltonian and the one describing the
lattice atoms do no longer commute, and the QME is only valid for a
large hopping with $J \gg E_p$.

\subsection{Sub- and superdecoherence and probing the BEC\label{Sec:Subandsuper}}

It has been predicted in Ref.~\cite{Palma-Proc-1996} that
decoherence effects caused by a qubit-bath coupling can be enhanced
(superdecoherence) or suppressed (subdecoherence) for certain
cases.\footnote{In our case it would be more appropriate to speak of
sub- and superdephasing. We will however stick to the conventions
introduced in Ref.~\cite{Palma-Proc-1996}.} Here we show that these
effects are indeed observable in a BEC for realistic experimental
parameters. The effect can moreover be used in order to probe such
BEC properties as the temperature with different sensitivity,
similar to the method introduced in Ref.~\cite{Bruderer-NJP-2006}.

Let us assume that the atoms have two internal states $\ket{0}$ and
$\ket{1}$ which couple to the BEC with different coupling constants
$\kappa_0 = 0$, $\kappa_1 = \kappa$. This case can be easily
generalised to a non-zero $\kappa_0$. The different coupling
strengths can for example be realised in a
$^{40}\mathrm{K}-$$^{87}\mathrm{Rb}$ mixture. The positions of the
Feshbach resonances for the scattering with the rubidium atoms
depends on the fine structure levels $\ket{F,m_F}$ of the potassium
atoms \cite{Ferlaino-PRA-2006}. By choosing the external magnetic
field close to one Feshbach resonance considerable differences
between the scattering lengths and hence the coupling constants for
different internal states can be achieved, and it is even possible
to tune one scattering length to zero.

Initially, the total state of the two atoms is described by the
density matrix $\hat \varrho_\mathbf{2}(0) = \sum_{ijkl = 0,1}
\varrho_{ijkl}(0) \ket{ij}\bra{kl}$. After evolving for a certain
time $t$ and tracing out the BEC the density matrix changes to $\hat
\varrho_\mathbf{2}(t) = \hat U_\mathrm{coh} \sum_{ijkl = 0,1}
\overline{\varrho}_{ijkl}(t) \ket{ij}\bra{kl} \hat
U_\mathrm{coh}^\dagger$, where $\hat U_\mathrm{coh}$ describes the
coherent evolution of the atoms induced by the off-site interaction
term, which for our case only changes the phase of the matrix
elements, but not their absolute value. The elements
$\overline{\varrho}_{ijkl}(t)$ are given by
$\overline{\varrho}_{ijkl}(t) = \varrho_{ijkl}(0)$ if $i = k$ and $j
= l$, $\overline{\varrho}_{ijkl}(t) = \Gamma_0 \varrho_{ijkl}(0)$ if
$i+j+k+l$ is an odd number, $\overline{\varrho}_{ijkl}(t) = \Gamma_-
\varrho_{ijkl}(0)$ for $\varrho_{0110}$ and $\varrho_{1001}$ and
$\overline{\varrho}_{ijkl}(t) = \Gamma_+ \varrho_{ijkl}(0)$ for
$\varrho_{1100}$ and $\varrho_{0011}$. The functions
$\Gamma_{+,0,-}$ are calculated analogously to the dephasing term
$\Gamma$ in Sec.~\ref{Sec:AnalytSol} and are given by
\begin{eqnarray} \fl
  \Gamma_0 = \exp\left( - \left.\sum_\qq\right.^\prime d_\qq
  \frac{1 - \cos(\omega_\qq t)}{(\hbar^2 \omega_\qq)^2} (2 N_\qq(T) +1)
  \right) \,,\\ \fl
   \Gamma_- = \exp\left( - \left.\sum_\qq\right.^\prime d_\qq
   [2 - 2 \cos(\qq (\br_\gamma - \br_\beta))]
  \frac{1 - \cos(\omega_\qq t)}{(\hbar^2 \omega_\qq)^2} (2 N_\qq(T) +1)
  \right) \,,\\ \fl
   \Gamma_+ = \exp\left( - \left.\sum_\qq\right.^\prime d_\qq
   [2 + 2 \cos(\qq (\br_\gamma - \br_\beta))]
  \frac{1 - \cos(\omega_\qq t)}{(\hbar^2 \omega_\qq)^2} (2 N_\qq(T) +1)
  \right) \,,
\end{eqnarray}
where $\br_\gamma$ and $\br_\beta$ denote the positions of the two
atoms.
\begin{figure}
\begin{center}
\includegraphics[width=12cm]{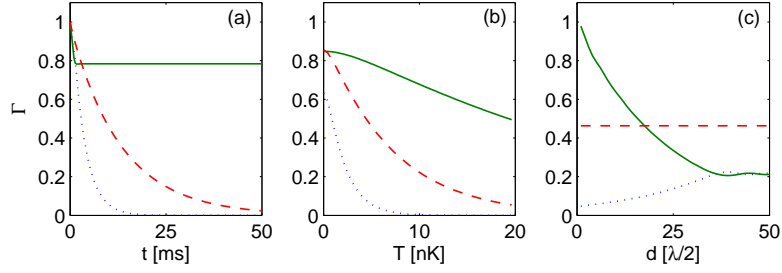}
\caption{The different behaviour of the dephasing functions
$\Gamma_-$ (solid line), $\Gamma_0$ (dashed line) and $\Gamma_+$
(dotted line) versus (a) time, (b) temperature, and (c) distance $d$
of the lattice atoms. Two $^{133}\mathrm{Cs}$ atoms were assumed in
an optical lattice with wavelength $\lambda = 790 \mathrm{nm}$ and a
depth of $V_0 = 40 E_\mathrm{R}$, where $E_\mathrm{R} = (2 \pi
\hbar)^2/2 m_l \lambda^2$. The one-dimensional BEC consists of
$^{87}\mathrm{Rb}$ atoms with a linear number density of $5\times
10^{6} /\mathrm{m}$. The couplings are given by $g = 4.5 \times
10^{-2} E_\mathrm{R} \lambda$ and $\kappa = 3.5 \times 10^{-2}
E_\mathrm{R} \lambda$. In (a), a temperature of $T = 5 \mathrm{nK}$
and a distance of $d = 5 \lambda/2$ were chosen, in (b) the same
distance and a time of $t = 10\mathrm{ms}$, in (c) $T= 5
\mathrm{nK}$ and $t = 10\mathrm{ms}$. For recent experiments of
RbCs-mixtures, see
Refs.~\cite{Tiesinga-PRA-2007,Anderlini-PRA-2005}.\label{Fig:DecayGammas}}
\end{center}
\end{figure}
Although the overall structure of these three decoherence terms
looks very similar they behave quite differently. Let us for the
moment assume that only such momenta $\qq$ contribute considerably
to the sum for which the condition $\qq ( \br_\gamma - \br_\beta)
\ll 1$ holds. This is typically the case for a temperature of $T
\approx 0$ and the atoms trapped in two neighbouring lattice sites.
Then, $\Gamma_- \approx 1$, which means that the decoherence for an
initial state $\ket{\psi} = (\ket{10} \pm \ket{01})/\sqrt{2}$ is
strongly suppressed, because the fluctuations of the condensate
happen on a length scale which is too large to resolve the distance
between the two atoms. In contrast, the decay of the off-diagonal
elements of the initial state $\ket{\psi} = (\ket{11} \pm
\ket{00})/\sqrt{2}$ is enhanced. These effects of sub- and
superdecoherence have been predicted in Ref.~\cite{Palma-Proc-1996}
for a general type of qubit-environment coupling and are similar to
the Dicke effect \cite{Dicke-PR-1954} well known in quantum optics.
However, in our case it is not the decay rate of the excited state
which is suppressed or enhanced. Instead, only the off-diagonal
elements of the density matrix are affected leaving the occupation
numbers unchanged.

As depicted in Fig.~\ref{Fig:DecayGammas}, these effects can indeed
be observed for realistic experimental parameters.
Figure~\ref{Fig:DecayGammas}(a) shows the time dependence of the
three functions. $\Gamma_-$ reaches its stationary state very
quickly, whereas the two other functions drop exponentially to zero.
The temperature dependence is shown in
Fig.~\ref{Fig:DecayGammas}(b). Since for higher temperatures phonons
with shorter wavelengths become more important, the condition $\qq (
\br_\gamma - \br_\beta) \ll 1$ is no longer fulfilled and $\Gamma_-$
decreases, however considerably slower than $\Gamma_+$. The
condition is also violated when the distance between the two atoms
is increased, as shown in Fig.~\ref{Fig:DecayGammas}(c).
Interestingly, the function $\Gamma_+$ shows exactly the opposite
behaviour: With increasing distance, the value of the function
increases as well, until it reaches the same value as $\Gamma_-$.

\subsection{Implications for the quantum register}

The dephasing investigated in the previous section has severe
implications on the performance of a quantum register. In order to
store the information two internal states $\ket{0'}$ and $\ket{1'}$
of the lattice atoms are needed, and we assume that these states do
not couple to the BEC, in which case there is also no decoherence
caused by the condensate coupling. However, the BEC can be used to
perform a two-qubit gate between two atoms submerged into it as
detailed in Ref.~\cite{Klein-PRA-2005}. In short, if an atom is in
state, say, $\ket{1'}$, it will be driven by a laser pulse to a
state $\ket{1}$ which couples to the condensate, whereas the state
$\ket{0'}$ is either unaffected or driven to a state $\ket{0}$ which
does not couple to the BEC. If both atoms involved are in state
$\ket{1}$ they can exchange BEC phonons which causes an additional
energy shift. In the previous section, this evolution was described
by the unitary $\hat U_\mathrm{coh}$, and by appropriately chosen
laser pulses this operator results in a controlled-phase gate.
However, due to the coupling the atoms also get entangled with the
BEC, which results in dephasing after tracing out the condensate.

The influence of the dephasing on the density matrix $\hat
\varrho_\mathbf{2}$ is suitably expressed by using Kraus operators
$E_j$. The super-operator $\Lambda$ giving the effect of the
dephasing (excluding the controlled-phase gate) can be expressed as
\begin{equation}
  \Lambda(\hat \varrho_\mathbf{2}) = \sum_{j = 1}^{d} E_j \hat
  \varrho_\mathbf{2} E_j^\dagger \,,
\end{equation}
where $d$ is the dimensionality of the quantum system, here $d=4$
for two qubits. Since the dephasing $\Lambda$ commutes with the
controlled-phase operation, the effect of the noisy gate is
described by $\Lambda_g(\hat \varrho_\mathbf{2}) = \hat U_{cZ}
\Lambda(\hat \varrho_\mathbf{2}) \hat U_{cZ}^\dagger = \Lambda(\hat
U_{cZ} \hat \varrho_\mathbf{2} \hat U_{cZ}^\dagger) $. The Kraus
operators for $\Lambda$ are given in~\ref{App:Kraus}, where we also
show how the average fidelity $ \langle F \rangle $ of the noisy
gate can be calculated using the explicit form of these operators.
We find
\begin{equation} \label{Eq:AvFid}
   \langle F \rangle =  \frac{1}{10} (4 + 4 \Gamma_0 + \Gamma_- +
   \Gamma_+) \,.
\end{equation}
From the definition of the dephasing terms it is evident that the
fidelity is worse for higher interactions $\kappa$ between the BEC
and the lattice atoms. On the other hand, a lower interaction
$\kappa$ means a lower interaction strength between the two atoms
and hence it takes a longer time to perform the gate. In the
three-dimensional case, by choosing an arbitrary low interaction
strength $\kappa$, the average gate fidelity can be brought
arbitrarily close to 1, however the time to perform the gate gets
arbitrarily long as well. Since there also exist other decoherence
mechanisms in cold atom systems, this will eventually decrease the
performance of the setup.

For concreteness, we consider the three-dimensional case where two
$^{133}\mathrm{Cs}$ atoms are placed in two neighbouring lattice
sites with wavelength $\lambda = 790\mathrm{nm}$. They are
surrounded by a $^{23}\mathrm{Na}$ BEC with a number density of
$(5\times 10^6)^3/\mathrm{m}^3$ and an interaction strength of $g =
1.5 \times 10^{-2} E_\mathrm{R} \lambda^3$. Here, $E_\mathrm{R} = (2
\pi \hbar)^2/2 m_\mathrm{l} \lambda^2$ is the recoil energy, with
$m_\mathrm{l}$ the mass of the lattice atoms and $\lambda$ the wave
length of the laser creating the lattice. The interspecies coupling
is given by $\kappa = 1.1 \times 10^{-2} E_\mathrm{R} \lambda^3$. In
the thermodynamic limit, for $i\neq j$ and $\xi \gg x_0$ the
mediated interaction is $V_{i,j} = \kappa^2 \xi \exp(-\sqrt{2}
|i-j|a/\xi)/\pi g \xi^3 |i-j| a$, where $a$ is the distance between
two neighbouring lattice sites. This leads to the (minimal) gate
time $t_g = \hbar \pi/V_{1,2} = 40 \mathrm{ms}$
\cite{Klein-PRA-2005}, where the gate fidelity is given by
$\left\langle F \right \rangle = 0.99$. For getting the last result,
we used the approximations $\Gamma_x \geq \exp(- c_x
\kappa^2/\sqrt{2} \pi^2 g^2 n_0 \xi^3)$, which hold in the
thermodynamic limit, $T=0$, $\xi \gg x_0$, and where $c_0 = 1$, $c_-
= 2$, and $c_+ = 4$.

For the case of two or one spatial dimensions, the interaction is in
general larger and thus the expected gate times shorter. However,
the fidelities decrease much faster than for three dimensions, and
all of our numerical tests showed that for the same gate times the
three-dimensional fidelities were always better than the ones
achieved in lower dimensions. Although this restricts the
applicability of this setup for quantum information purposes, the
scheme can still be used for probing the interaction strength which
is mediated by the condensate, simply by measuring the phase
differences for varying interaction times.


\section{Clustering of the lattice atoms \label{Sec:Cluster}}

If the lattice is not uniformly filled with atoms but the filling
ranges somewhere between zero and one atom per lattice site, the
mediated interaction has an important impact on the spatial
distribution of the atoms. For low enough temperatures, it will lead
to atom clusters. To observe the clustering of the atoms we have to
allow for a weak hopping $J \ll |V_{1,2}|,|U|$,\footnote{In the
assumed homogeneous setup the value $V_{1,2}$ corresponds to the
mediated interaction between two nearest neighbours and $V_{1,1}$ to
the mediated on-site interaction.}, because otherwise the atom
distribution remains stationary. We can assume that the perturbation
due to the hopping does not change the interaction potential
$V_{i,j}$ derived from the exact solution, which is in agreement
with our earlier findings~\cite{Bruderer-PRA-2007}. The hopping
energy can furthermore be neglected compared to the interaction
potential, such that the weak hopping leads only to a re-arrangement
of the atoms, which is similar to the treatment of ad-atoms on
crystal surfaces~\cite{Jensen-RMP-1999}. The Hamiltonian for this
situation is given by
\begin{equation}
  \hat H_\mathrm{cl} = \frac{U}{2} \sum_{j} \hat n_j ( \hat n_j -1) -
  \sum_{i,j} V_{i,j} \hat n_i \hat n_j \,.
\end{equation}

\begin{figure}
\begin{center}
\includegraphics[width=14cm]{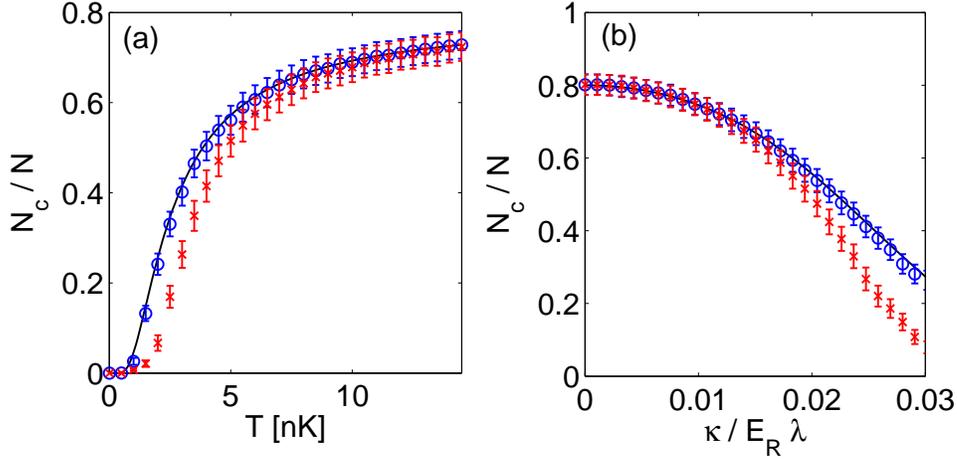}
\caption{Normalised cluster number $N_\mathrm{c}/N$ versus (a)
temperature of the one-dimensional lattice gas and (b) coupling to
the BEC. The solid line shows the analytical estimate given in
Eq.~(\ref{Eq:Numclust1D}), results from numerical calculations only
taking the nearest neighbour interaction into account are marked by
`o' and numerical results including the full interaction potential
are marked by `x'. The algorithm was run for $10^8$ steps after
which an equilibrium state is reached. Then, another $5\times 10^7$
steps were done and after every 1000 steps the number of clusters
was calculated. Error bars show one standard deviation derived from
averaging over those results. The lattice with wavelength $\lambda =
790\mathrm{nm}$ consists of $M = 800$ lattice sites filled with
$N=160$ $^{41}\mathrm{K}$ atoms. The surrounding BEC was assumed to
consist of $^{87}\mathrm{Rb}$ atoms with a linear number density of
$n = 5 \times 10^6/\mathrm{m}$ and a coupling $g/E_\mathrm{R}
\lambda  = 1.1 \times 10^{-2}$. In (a), the interspecies coupling is
given by $\kappa/E_\mathrm{R} \lambda  = 2.5 \times 10^{-2}$,
whereas in (b) the temperature is fixed to $T=3\mathrm{nK}$.
Periodic boundary conditions were assumed.
\label{Fig:Clusternumber}}
\end{center}
\end{figure}

\begin{figure}
\begin{center}
\includegraphics[width=14cm]{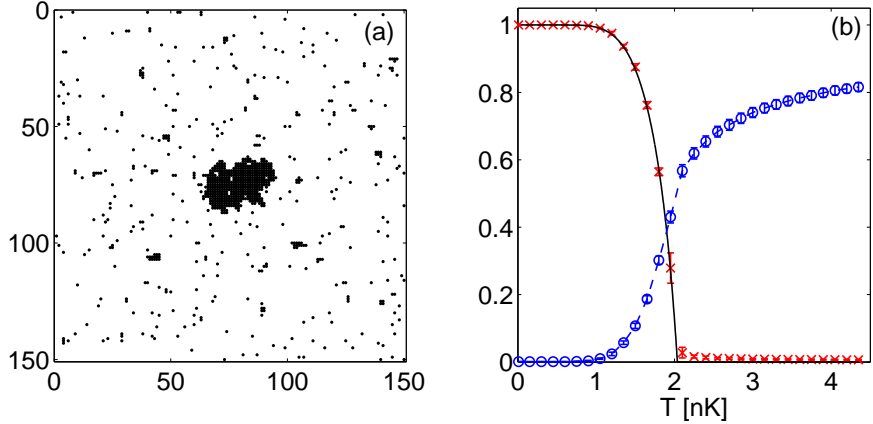}
\caption{(a) Distribution of the atoms in a homogenous
two-dimensional lattice with periodic boundary conditions and a
temperature of $T = 1.8 \mathrm{nK}$. A macroscopic island is
clearly visible. (b) Normalised size of the largest cluster $N_I/N$
versus temperature of the lattice gas for a two-dimensional system
(marked by `x') and normalised number of clusters $N_c/N$ (marked by
`o'). The solid line shows the analytical expression given in
Eq.~(\ref{Eq:IslandSize2D}), the dashed line is a guide to the eye.
Error bars show one standard deviation derived from averaging over
several runs. The lattice with wavelength $\lambda = 790\mathrm{nm}$
consists of $M = 150 \times 150$ lattice sites filled with 900
$^{41}\mathrm{K}$ atoms. The surrounding BEC was assumed to consist
of $^{87}\mathrm{Rb}$ atoms with a number density of $n = 25 \times
10^{12}/\mathrm{m^2}$ and a coupling $g/E_\mathrm{R} \lambda^2 = 5.1
\times 10^{-3}$. The interspecies coupling is given by
$\kappa/E_\mathrm{R} \lambda^2  = 1.87 \times 10^{-2}$.
\label{Fig:Clusternumber_2D}}
\end{center}
\end{figure}

The mediated on-site interaction $V_{1,1}$, which in our case is
always attractive, cf.~equation (\ref{Eq:Vlls}), might
overcompensate the on-site interaction $U$, i.e., $U + V_{1,1} < 0$.
If this happens, all the atoms can aggregate in a single lattice
site, which will inevitably lead to three-body losses. We therefore
assume that the repulsive on-site interaction $U$ is large enough
such that states with more than one atom in a single lattice site
can be neglected, i.e., $U + V_{1,1} \gg J$. For the parameters used
below (see captions of Figs.~\ref{Fig:Clusternumber} and
\ref{Fig:Clusternumber_2D}) the mediated on-site interaction is
given by $V_{1,1}= -0.03 E_\mathrm{R}$ in one dimension and
$V_{1,1}= -0.08 E_\mathrm{R}$ in two. Since we require a deep
lattice for the hopping $J$ to be small, a sufficiently high $U$ can
easily be achieved. For a reasonably deep lattice of $15
E_\mathrm{R}$ and perpendicular confinement of $35 E_\mathrm{R}$,
the interaction strength is on the order of $U \approx 0.4
E_\mathrm{R}$ for a one-dimensional lattice and $U \approx 0.3
E_\mathrm{R}$ for two dimensions and thus sufficiently high to
overcome the induced on-site attractive interaction. The repulsive
interaction strength corresponds to a temperature of $U/k_\mathrm{B}
= 150 \mathrm{nK}$ in one dimension ($U/k_\mathrm{B} = 130
\mathrm{nK}$ in two dimensions), such that neglecting states with
more than one atom per lattice site is well-justified for the
temperature regime considered, see below. With these assumptions,
the ground state for the lattice system consists of a cluster where
all the atoms are located in neighbouring lattice sites. We note
that this also holds if the lattice is loaded with spin-polarised
fermions, in which case the Pauli exclusion principle ensures that
there is never more than one atom in each lattice site. The
restriction to a maximum of one atom per lattice site and ignoring
interactions beyond nearest neighbours makes it also possible to map
the system to an Ising model by using the correspondence $\hat n_j =
(1 + \hat s_j)/2$, which gives
\begin{equation}
  \hat H_\mathrm{cl} = - \frac{V_{1,2}}{4} \sum_{\langle i,j\rangle}
  \hat s_i \hat s_j - V_{1,2} \sum_j \hat s_j \,,
\end{equation}
where a constant term has been neglected. We will make use of this
correspondence shortly when we compare our results to analytical
ones found for the Ising model in two dimensions.

In an experiment, the lattice atoms will have a finite temperature
$T$, which leads to a breaking up of the ground state into smaller
clusters, reflecting the increased average energy. It is therefore
important to investigate at which temperatures the clustering can be
observed. To this end, we simulated the lattice system using the
well-established Metropolis
algorithm~\cite{Metropolis-JCP-1953,Binder-1986}, which is often
used in the simulation of classical lattice spin and lattice gas
models when the kinetic energy can be neglected compared to the
interaction energies. For a one-dimensional lattice, the averaged
number of clusters\footnote{Note that a single atom without nearest
neighbours is also considered to constitute a cluster.} for
different temperatures is shown in Fig.~\ref{Fig:Clusternumber}(a).
For temperatures above $T = 7 \mathrm{nK}$ the number of clusters
does not change very much with the temperature and roughly 110
clusters exist, which shows that most of the atoms do not have
nearest neighbours. A drastic reduction of the cluster number only
occurs for $T < 7 \mathrm{nK}$, such that the number of clusters
finally reaches $N_\mathrm{c} = 1$ for $T \approx 0$. A method to
achieve such low temperatures of below 7\,nK has been proposed,
where the surrounding BEC is first used for dark-state cooling
\cite{Griessner-PRL-2006,Griessner-NJP-2007}.

The numerical data in Fig.~\ref{Fig:Clusternumber} is compared to an
analytical result derived in the thermodynamic limit, where the
number of lattice sites $M$ goes to infinity whilst keeping the
filling fraction $N/M$ constant, and only taking nearest-neighbour
interactions into account \cite{Yilmaz-PRE-2005}. In this case, the
normalised number of clusters $N_\mathrm{c}/N$ is given by
\begin{equation}\label{Eq:Numclust1D}
  \frac{N_\mathrm{c}}{N} = \frac{M}{N} \frac{\sqrt{1 + 4 (M-N)N
  \left[\exp(2|V_{1,2}|/k_\mathrm{B} T)-1\right]/M^2}-1}{2[\exp(2|V_{1,2}|/k_\mathrm{B}
  T)-1]} \,,
\end{equation}
where $k_\mathrm{B}$ is Boltzmann's constant. Our calculations show
that the numerical results taking only the nearest neighbour term of
the interaction into account is in excellent agreement with the
result for the thermodynamic limit. We furthermore observe that the
number of clusters for the full interaction potential is for low
temperatures considerably smaller than for the truncated one, which
indicates that the interaction terms beyond nearest neighbours make
the clusters more stable. This gets also apparent when the
temperature is fixed and the number of clusters is calculated for
different coupling strengths $\kappa$, as shown in
Fig.~\ref{Fig:Clusternumber}(b). For small $\kappa$ the interaction
beyond nearest neighbours is still quite weak and does not change
the number of clusters significantly, whereas for stronger coupling
$\kappa$, interaction terms beyond nearest neighbours are
considerable, reflected in a smaller number of clusters or,
equivalently, in a larger average cluster size.

The situation gets more interesting if we consider a two-dimensional
lattice. For $T \approx 0$, the atoms aggregate in an ``island'' and
for increasing temperature parts of this island break away, see
Fig.~\ref{Fig:Clusternumber_2D}(a). We have calculated the size of
the largest cluster (i.e.~the number $N_I$ of atoms constituting the
island) numerically, only taking nearest neighbour interactions into
account. This will give a lower bound for the temperature below
which the formation of the island can be observed. The results are
shown in Fig.~\ref{Fig:Clusternumber_2D}(b). For very low
temperatures, all the atoms are contained in a single cluster. This
largest cluster gets smaller with increasing temperature and shows a
pronounced transition at a temperature of $T \approx 2 \mathrm{nK}$,
above which the system mainly consists out of many small clusters.
It has been shown in Ref.~\cite{Pleimling-JPA-2000}, that the
normalised size $N_I/N$ of the island in the thermodynamic limit and
only taking nearest neighbour interactions into accout behaves as
\begin{equation}\label{Eq:IslandSize2D}
  \frac{N_I}{  N} = \frac{(1 + N_0) (N_0 - N_{r})}{2 N_0 (1-N_r)}
  \,.
\end{equation}
Here, $N_0 = \{ 1 - [\sinh(2 J_\mathrm{s}/k_\mathrm{B} T)]^{-4}
\}^{1/8}$, and $J_\mathrm{s} = |V_{1,2}|/4$. For this formula to be
valid we further require a filling of $N/M \leq 1/2$, such that $N_r
= 1 - 2 N/M \geq 0$, and above the transition temperature $T_I$
where $N_I/M$ hits zero for the first time, the function has to be
set to zero, since for this temperature the normalised size of the
largest island vanishes in the thermodynamic limit. Our Monte-Carlo
results are in excellent agreement with Eq.~(\ref{Eq:IslandSize2D}),
deviations above the transition temperature are due to finite size
effects. We note that a higher transition temperature $T_I$ can be
achieved by increasing the interaction strength $V_{i,j}$ or by
loading more atoms into the lattice. We find $k_\mathrm{B} T_I = 2
J_\mathrm{s} / \mathrm{arsinh} \left[1/\sqrt[4]{1-N_r^8} \right]$,
which yields for the chosen values (see caption of
Fig.~\ref{Fig:Clusternumber_2D}) and $N_r \to 0$ a value of $T_I =
2.34 \mathrm{nK}$.


\section{Transport properties of the impurity atoms \label{Sec:QTrans}}

\begin{figure}
\begin{center}
\includegraphics[width=9cm]{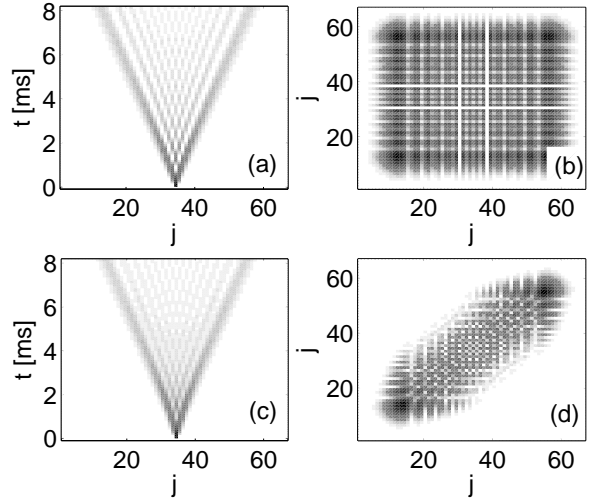}
\caption{(a) and (c) time evolution of the atom density. (b) and (d)
graphical representation of the absolute values of the density
matrix for the final time step $t = 8.2\mathrm{ms}$ shown in (a) and
(c). Darker colour corresponds to a higher numerical value. For the
examples shown we have assumed a $^{41}\mathrm{K}$ atom in an
optical lattice with wave length $\lambda = 790\mathrm{nm}$, and a
hopping of $J = 0.03 E_\mathrm{R}$. The BEC consists of
$^{87}\mathrm{Rb}$ atoms with a density of $n_0 = 5 \times 10^6
/\mathrm{m}$ and a coupling of $g/E_\mathrm{R} \lambda = 1.1 \times
10^{-2}$. The temperature is given by $T = 100 \mathrm{nK}$. The
interspecies coupling for case (a) and (b) was assumed to be $\kappa
= 0$, whereas for (c) and (d) it was $\kappa/E_\mathrm{R} \lambda =
1.94 \times 10^{-2}$. For recent experiments on these mixtures, see
Refs.~\cite{Modugno-PRL-2002,Catani-2007}. \label{Fig:spreading}}
\end{center}
\end{figure}

In this section, we use the QME to investigate the behaviour of the
lattice system for a hopping $J \gg E_p$. In general, an analytical
solution of the QME is no longer possible, and the dynamics has to
be calculated numerically. Let us consider the simple case of one
atom initially localised in a single lattice site. With vanishing
coupling to the condensate, the atom will spread across the lattice
coherently in a wavelike motion, as shown in
Fig.~\ref{Fig:spreading}(a). The interference fringes between the
two wave fronts are clearly visible. The coherent nature of the
evolution is also reflected in the density matrix, where all
off-diagonal elements have their maximum possible value, compare
Fig.~\ref{Fig:spreading}(b). The situation changes when the coupling
to the condensate is increased, as shown in
Fig.~\ref{Fig:spreading}(c). The wave-fronts still exist, however
the region in between no longer exhibits clear interference
patterns, which are washed out instead. This implies that the
coherent nature of the evolution is impaired, which is also
supported by the density matrix, see Fig.~\ref{Fig:spreading}(d).
The off-diagonal elements are clearly suppressed, which confirms the
incoherent character.

\begin{figure}
\begin{center}
\includegraphics[width=9cm]{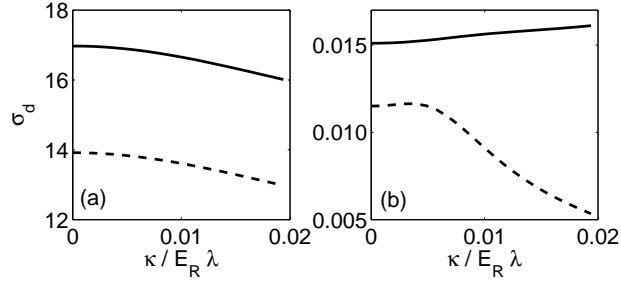}
\caption{(a) Standard deviation of the density distribution after an
evolution time of $t = 6.7\mathrm{ms}$ (dashed line) and $t =
8.2\mathrm{ms}$ (solid line). (b) Average density $\bar{p}$ (solid
line) and standard deviation $p_\mathrm{d}$ (dashed line). Apart
from the interspecies coupling $\kappa$, all the parameters are as
in Fig.~\ref{Fig:spreading}. \label{Fig:cross}}
\end{center}
\end{figure}

The motion of the atom in the lattice can be explained by the
coexistence of wavelike, coherent evolution responsible for the two
wave-fronts, and a diffusive, incoherent evolution. The diffusive
motion stems from the decay of the off-site elements in the density
matrix, which destroys the memory of the atom and causes it to
perform a random walk, leading to the washing out of the
interference effects between the two wave fronts. To investigate the
time evolution more quantitatively, we calculate the standard
deviation $\sigma_\diff$ of the atom density distribution at a time
$t$, given by $\sigma_\diff = \sum_j p_j (j - j_0)^2$, where $j$
labels the lattice sites, $p_j$ is the probability of finding the
atom at lattice site $j$, and $j_0$ is the initial lattice site of
the atom at $t=0$. Figure~\ref{Fig:cross}(a) shows that for an
increasing coupling to the condensate this standard deviation
$\sigma_\diff$ decreases. This stems from the fact that the hopping
is reduced due to the coupling between the lattice atoms and the
phonons of the BEC, which has also been observed in
Ref.~\cite{Bruderer-PRA-2007}.

Taking one standard deviation to either side of $j_0$ defines a
suitable interval $I$ between the two wave fronts to investigate the
diffusive character of the motion. For this interval, we calculate
the average atom density $\bar{p} = \sum_{j \in I} p_j / 2
\sigma_\diff$ and the standard deviation of the density distribution
$p_\diff = \sum_{j \in I} (p_j - \bar{p})^2/2 \sigma_\diff$. For the
coherent case, we expect that this standard deviation is on the
order of the average density $\bar{p}$ due to the interference
patterns, whereas for the incoherent case the standard deviation
should be much lower than the average density due to the more
homogenous spread of the atom between the two wave fronts. These
expectations are confirmed by our findings shown in
Fig.~\ref{Fig:cross}(b). We observe that the average density between
the two wave fronts $\bar p$ moderately increases with increasing
interspecies coupling. The standard deviation of this density
$p_\diff$ stays first approximately constant with increasing
$\kappa$, indicating a regime where the evolution can still be
considered to be coherent, but then for $\kappa > 5\times 10^{-3}
E_\mathrm{R} \lambda$ drops off considerably, caused by the
increased loss of coherence.

The QME describes the lattice atoms well as long as only dephasing
has to be taken into account and energy exchange with the BEC can be
neglected. This is consistent with the approximations we have used
to derive the QME: It was assumed that $J a /\hbar \ll c$, which
means that the atoms move slower than the critical velocity of the
BEC. Then, according to the Landau criterion for superfluidity, no
energy exchange is possible with a single phonon process, and higher
order phonon processes are not included into the QME.

\begin{figure}
\begin{center}
\includegraphics[width=9cm]{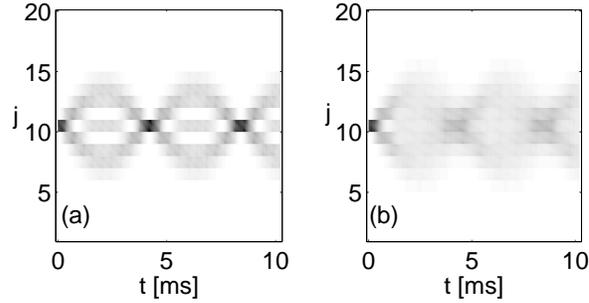}
\caption{Bloch oscillations in an optical lattice submerged into a
BEC of $^{87}\mathrm{Rb}$ atoms with a density of $n_0 = 5 \times
10^6 /\mathrm{m}$, a coupling of $g/E_\mathrm{R} \lambda = 1.1
\times 10^{-2}$ and a temperature of $T = 75 \mathrm{nK}$. The
$^{41}\mathrm{K}$ atom is trapped in a lattice with wavelength
$\lambda = 790\mathrm{nm}$ and has a hopping rate of $J = 0.03
E_\mathrm{R}$. The strength of the Stark potential is given by
$K/\hbar = 1.5 \mathrm{kHz}$. The interspecies coupling is given by
(a) $\kappa = 0$ and (b) $\kappa/E_\mathrm{R} \lambda = 1.6\times
10^{-2}$. \label{Fig:Bloch_QME}}
\end{center}
\end{figure}

This gets especially apparent when we try to describe the decay of
Bloch oscillations with the QME. Let us assume a Stark potential
$\hat H_s = K \sum_j j \hat n_j$ is applied to the one-dimensional
optical lattice. For one particle initially located in a single
lattice site and not coupling to the BEC, we expect breathing
oscillations with frequency $\omega_B = K/\hbar$ and width $4 J_a |
\sin(\omega_B t/2)|/K $ \cite{Hartmann-NJP-2004}. This is shown in
Fig.~\ref{Fig:Bloch_QME}(a). When the atom couples to the BEC, it
can dissipate energy into the condensate modes and one expects that
the atom can ``fall down'' the tilted lattice, causing a net
current. When using the QME to describe the dynamics this is,
however, not the case. Instead, the Bloch oscillations only get
washed out, but the mean position of the atom remains unchanged, as
shown in Fig.~\ref{Fig:Bloch_QME}(b). This is in agreement with our
expectations and with other attempts to describe the system with a
simple Master Equation approach~\cite{Ponomarev-PRL-2006}. An
alternative way based on a generalised master equation, which
includes the coupling of phonons to the hopping term, has been
devised to describe the occurrence of an atomic net current. Details
will be given elsewhere \cite{Bruderer-arXiv-2007}.

\begin{figure}
\begin{center}
  \includegraphics[width=9cm]{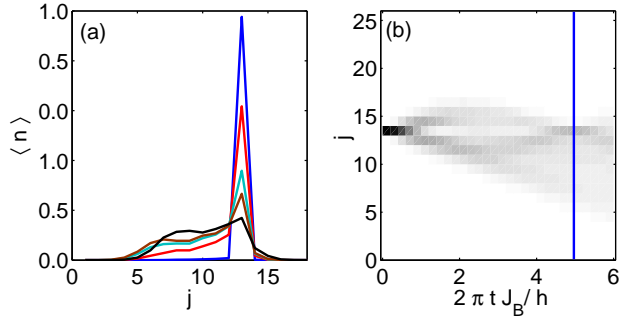}
  \caption{(a) Decay of Bloch oscillations and occurrence of a net current. Shown
  are the density distributions of one atom initially localised
  in lattice site 13 after a time evolution of $t = 2 \pi/\omega_B$
  for different interaction strengths ranging from $U_i =\{ 0.1;0.5;0.8;1;2  \} J_b$
  (top to bottom line in lattice site 13).
  The dynamics of the lattice atom and
  the BEC consisting of 25 atoms was calculated
  in an optical lattice with 25 lattice sites using the TEBD algorithm.
  We used $J_a = J_b$, $U_b = 0.5 J_b$, $K = 1.25 J_b$, $T=0$. (b) Time
  evolution of the density
  distribution for one lattice atom with $U_i = J_b$. The coexistence of
  Bloch oscillations and a current is visible. The vertical line
  indicates the time where the densities of (a) are taken.
       \label{Fig:Bloch}}
\end{center}
\end{figure}

Another way to observe the net current is given by solving the
complete dynamics of the atom and the surrounding BEC numerically.
For this, the BEC has to be discretised, which is equivalent to
assuming that it is trapped in an optical lattice with appropriately
chosen hopping and interaction constants \cite{Schmidt-PRA-2005}.
The Hamiltonian describing the system is then given by
\begin{equation}
\fl
  \hat H = \hat H_\mathrm{L} -J_b \sum_{\langle i,j \rangle} \hat
\beta_i^\dagger \hat \beta_j + \frac{U_b}{2} \sum_j \hat
\beta^\dagger_j \hat \beta^\dagger_j \hat \beta_j \hat \beta_j + U_i
\sum_j \hat \beta^\dagger_j \hat \beta_j \hat a_j^\dagger \hat a_j
\,,
\end{equation}
with $\hat \beta_j^{\dagger}$ $(\hat \beta_j)$ an operator that
creates (annihilates) a boson in lattice site $j$, $U_b$ the on-site
interaction strength and $J_b$ the hopping matrix element of the BEC
atoms, and $U_i$ the interspecies interaction strength. The time
evolution of this Hamiltonian including the BEC and a single atom in
the lattice is performed numerically by using the time evolving
block decimation (TEBD)
algorithm~\cite{Vidal-PRL-2004,Verstraete-PRL-2004,Clark-PRA-2004}.
The resulting density distribution of the lattice atom initially
placed in a single lattice site and evolved for one Bloch period is
shown in Fig.~\ref{Fig:Bloch}(a). For a weak interaction strength
$U_i \ll J_b$ the surrounding BEC atoms hardly affect the dynamics
of the lattice atom and after one cycle it ends up in its original
lattice site again. For increased interaction the lattice atom is
able to dissipate more and more energy to the BEC and to drift
towards lattice sites with a lower potential energy, which leads to
a net current. This is illustrated in Fig.~\ref{Fig:Bloch}(b), where
a competition between Bloch oscillation and net current is clearly
visible. The current can be measured either directly or after a time
of flight expansion of the lattice atoms as detailed in
Ref.~\cite{Schulte-2007}.


\section{Conclusion}

In the present paper we have investigated the behaviour of an
optical lattice system immersed into a BEC. To this end, we have
derived a quantum master equation which describes the time evolution
of the lattice atoms. For the case of fixed impurities, the lattice
system represents a quantum register and the internal states of the
atoms represent the states of the qubit. We have derived an exact
solution for this system which is reproduced by the QME. We found
that an interaction between the lattice atoms is mediated by the
condensate and that the coupling to the phonon modes of the BEC
causes a dephasing of the lattice atoms, i.e., of the qubits they
represent. This dephasing can be used to probe the properties of the
BEC with different sensitivity.

The interaction between the lattice atoms leads to a clustering
process at low enough temperatures. In contrast to on-site clusters,
which are prone to three-body losses, the clustering in our case is
caused by the off-site interaction terms. We have simulated the
clustering process using the Metropolis algorithm and compared our
results with analytical ones originally found for spin systems or
ad-atoms on a crystal surface. Our findings indicate that for
realistic experimental parameters the clustering process occurs at
temperatures of about 5nK and is observable with near-future
experimental techniques.

We also investigated the transport properties of the lattice atoms
subject to the BEC coupling. We found that the lattice atoms undergo
a crossover from coherent to incoherent evolution when the
interaction to the condensate is increased. This process is
indicated by a washing out of the typical interference fringes in
the density distribution and by vanishing off-diagonal elements in
the one-particle density matrix. However, due to the approximations
necessary to derive the QME, it does not include dissipation of
energy to the BEC. This process is a vital effect to describe the
decay of Bloch oscillations in a tilted lattice. More work is
necessary to include this effect into a master equation description.

\ack{A.K. thanks Bernd Schmidt and Michael Fleisch\-hauer for
fruitful discussions. This work was supported by the UK EPSRC
through QIP IRC (GR/S82176/01) and EuroQUAM project EP/E041612/1,
the EU through the STREP project OLAQUI, the Berrow Scholarship
(M.B.), and the Keble Association (A.K.).}

\appendix

\section{Derivation of the QME \label{App:QME}}

The details of deriving the QME are given in this Appendix. Our
starting point is the Liouville-von Neumann equation for the total
density operator $\hat \varrho(t)$,
\begin{equation}
  \ii \hbar \partial_t \hat \varrho(t) = [ \hat H, \hat \varrho (t)] \,.
\end{equation}
For a general operator $\hat O$ in the Schr{\"o}dinger picture we
define the corresponding operator $\tilde O(t)$ in the interaction
picture by
\begin{equation}
\tilde O (t) = \eh^{\ii (\hat H_B + \hat H_L )t/\hbar } \, \hat O
  \, \eh^{- \ii (\hat H_B + \hat H_L )t /\hbar} \,.
\end{equation}
With this, the Liouville-von Neumann equation changes to
\begin{equation}
  \ii \hbar \partial_t \tilde \varrho(t) = [ \tilde H_I(t), \tilde \varrho(t)] \,.
\end{equation}

If there was no interaction between the lattice atoms and the BEC
before a time $t_0 = 0$, it is reasonable to assume that the density
operator at time $t_0$ is described by $\hat \varrho(0) = \hat
\varrho_\mathrm{L}(0) \otimes \hat \varrho_\mathrm{B}(0)$, i.e., at
time $t_0$ there are no correlations present between the BEC and the
lattice atoms. Here, $\hat \varrho_\mathrm{L}(0)$ and $\hat
\varrho_\mathrm{B}(0)$ are the initial density operators of the
lattice system and the BEC, respectively. We furthermore assume that
at $t_0$ the BEC is in a thermal state with temperature $T$.
Integrating the Liouville-von Neumann equation, substituting it into
itself and using the definition of $\tilde H_I(t)$, we find after
applying the Born approximation and taking the trace over the BEC
variables \cite{Breuer-2002}
\begin{eqnarray}
  \nonumber\fl
  \partial_t \tilde \varrho_L(t) = - \frac{1}{\hbar^2} \int_0^t \!
  \diff t' \,& \left.\sum_\qq\right.^\prime \sum_{l,l'} \{   [ \tilde n_l(t)
  \tilde n_{l'}(t') \tilde \varrho_L(t') - \tilde n_{l'}(t') \tilde
  \varrho_L(t') \tilde n_l(t) ] C_{\qq,l,l'}(t-t')  \\  \fl
  & + [ \tilde \varrho_L(t') \tilde n_{l'}(t') \tilde n_l(t)
    - \tilde n_{l}(t) \tilde
  \varrho_L(t') \tilde n_{l'}(t') ] C^\ast_{\qq,l,l'}(t-t')
  \} \,,
\end{eqnarray}
with
\begin{equation}\fl
  C_{\qq,l,l'}(t-t') = F_{\qq,l} F^\ast_{\qq,l'} (1 + N_\qq(T))\, \eh^{- \ii \omega_\qq (t-t')}
  +  F^\ast_{\qq,l} F_{\qq,l'} N_\qq(T) \,\eh^{ \ii \omega_\qq
  (t-t')} \,.
\end{equation}
Under the
assumption that the sound velocity of the condensate, $ c \sim
\sqrt{g n_0/m_b}\,$, is larger than the typical hopping speed of the
atoms, $c \gg J a/\hbar$, the dynamics of the BEC is much faster
than the typical dynamics of the lattice atoms and we perform the
Markov-approximation by replacing $\tilde \varrho_L(t') \rightarrow
\tilde \varrho_L(t) $ and $\tilde n_l(t') \rightarrow \tilde
n_l(t)$. After calculating the time integral and transforming back
into the Schr{\"o}dinger picture we finally arrive at
Eq.~(\ref{Eq:QME}).

\section{Kraus operators for the two-qubit dephasing \label{App:Kraus}}
Here, we give explicit formulas for the Kraus operators describing
the operator $\Lambda$ given by $\Lambda(\hat \varrho_\mathbf{2}) =
\hat U_{cZ}^\dagger \Lambda_g(\hat \varrho_\mathbf{2}) \hat U_{cZ}$.
We find
\begin{eqnarray}
  E_1 = \sqrt{(1+2\Gamma_0 + \Gamma_+)/4}\, \mathbf{1}_4 \,, \\
  E_2 = \sqrt{(1-2\Gamma_0 + \Gamma_+)/4}\,  \sigma_z \otimes \sigma_z
  \,, \\
  E_3 = \sqrt{(1-\Gamma_-)/4}\,  \sigma_z \otimes \mathbf{1}_2 \,, \\
  E_4 = \sqrt{(1-\Gamma_-)/4} \,  \mathbf{1}_2 \otimes \sigma_z \,,
  \\
  E_5 = \sqrt{(\Gamma_- - \Gamma_+)/4}\, \mathrm{diag}(-1,1,1,1) \,,
  \\
  E_6 = \sqrt{(\Gamma_- - \Gamma_+)/4}\, \mathrm{diag}(1,1,1,-1) \,,
\end{eqnarray}
and all other Kraus operators are zero. Here, $\mathbf{1}_n$ is the
$n \times n$ unit matrix, $\sigma_z$ is the Pauli $z$ matrix, and
$\mathrm{diag}$ is a diagonal matrix with the entries on the
diagonal given as the argument.

With the knowledge of the Kraus operators the average fidelity
$\langle F \rangle $ of the gate is calculated as follows. We first
note that the fidelity $F$ of the noisy operation $\Lambda_g$ with
respect to the perfect (i.e., noise-free) operation $\hat U_{cZ}$ on
a special input state $\ket{\psi}$ is defined by
\begin{equation}
  F(\psi) =
  \bra{\psi}{\hat U^{\dagger}_{cZ} \, \Lambda_g\{\ket{\psi} \bra{\psi}\} \,
  \hat U_{cZ}}\ket{\psi}\,.
\end{equation}
By taking the average over all possible input states the overall
performance of the gate operation can be estimated. It has been
shown \cite{Dankert-2005,Pedersen-2007} that the average fidelity is
given by
\begin{equation}
  \langle F \rangle = \int_{S^{2d-2}}
  F(\psi)\,\mathrm{d}\psi = \frac{1}{d(d+1)}\left(\sum_{j=1}^{d^2}
|\tr(E_j)|^2 + d\right) \,,
\end{equation}
where the integral is taken over the unit sphere $S^{2 d -2}$
embedded in $2d-1$-dimensional real space, which is isomorphic to
the $d$-dimensional complex space after eliminating a global phase,
and $\diff \psi$ is the normalised measure over the sphere, also
know as Haar measure. With this formula and the explicit form of the
Kraus operators the calculation of the average fidelity
Eq.~(\ref{Eq:AvFid}) is straight forward.

We note that the Kraus operators for a coupling to independent
reservoirs are given by replacing $\Gamma_- = \Gamma_+ =
(\Gamma_0)^2$. This gives an average fidelity of $\langle
F_{\mathrm{ind}} \rangle = (2 + 2 \Gamma_0 + (\Gamma_0)^2)/5$. Hence
the coupling to the BEC reservoir is worse than the one to
independent reservoirs if $\Gamma_+ + \Gamma_- < 2 (\Gamma_0)^2$.

\section*{References}

\bibliographystyle{priv_bib_style}
\bibliography{LatticeBEC_long}

\end{document}